\documentclass[aps,notitlepage,12pt,showpacs,showkeys,tightenlines]{revtex4}

\usepackage{amsmath}
\usepackage{amssymb,amsfonts}
\usepackage{bm}
\usepackage[mathcal]{euscript}
\usepackage{graphicx}
\usepackage{psfrag}
\usepackage{subfigure}

\newcommand{\ie}{\textit{i.e.}}
\newcommand{\etal}{\textit{et~al.}}

\newcommand{\mathnotation}[2]{\newcommand{#1}{\ensuremath{#2}}}

\newcommand{\transp}[1]{{#1}^T}			
\newcommand{\Order}[1]{\ensuremath{\mathrm{O}\!\l(#1\r)}}

\newcommand{\goodgap}{%
	\hspace{\subfigtopskip}%
	\hspace{\subfigbottomskip}}

\renewcommand{\l}{\left}			
\renewcommand{\r}{\right}			
\mathnotation{\pd}{\partial}			
\mathnotation{\ee}{{\mathrm e}}			
\mathnotation{\ldef}{\mathrel{\raisebox{.069ex}{:}\!\!=}}
\mathnotation{\rdef}{\mathrel{=\!\!\raisebox{.069ex}{:}}}
\mathnotation{\dint}{\,{\mathrm{d}}}		

\mathnotation{\grad}{\nabla}			
\renewcommand{\div}{\grad\cdot}			
\mathnotation{\curl}{\grad\times}		
\mathnotation{\lapl}{\nabla^2}			

\renewcommand{\time}{t}				
\mathnotation{\xc}{x}				
\mathnotation{\xv}{{\bm{\xc}}}			
\mathnotation{\barxv}{\overline\xv}		
\mathnotation{\velc}{v}				
\mathnotation{\velv}{{\bm{\velc}}}		
\mathnotation{\flow}{\Phi}			
\mathnotation{\Mc}{M}				
\mathnotation{\Mt}{\mathbb{\Mc}}		
\mathnotation{\sdim}{n}				

\mathnotation{\edir}{e}				
\mathnotation{\edirv}{\mathbf{\edir}}		
\mathnotation{\ediru}{\hat{\edir}}		
\mathnotation{\ediruv}{\hat{\mathbf{\edir}}}	
\mathnotation{\ediruinf}{\ediru^\infty}		
\mathnotation{\ediruvinf}{\ediruv^\infty}	
\mathnotation{\edirt}{\tilde{\edir}}
\mathnotation{\edirtv}{\tilde{\mathbf{\edirv}}}
\mathnotation{\udir}{{\mathrm{u}}}		
\mathnotation{\udirv}{\mathbf{u}}		
\mathnotation{\udiru}{\hat{\udir}}		
\mathnotation{\udiruv}{\hat{\udirv}}		
\mathnotation{\udiruinf}{\udiru^\infty}		
\mathnotation{\udiruvinf}{\udiruv^\infty}	
\mathnotation{\udirt}{\tilde{\udir}}
\mathnotation{\udirtv}{\tilde{\udirv}}

\mathnotation{\lac}{a}				
\mathnotation{\lav}{{\bm{\lac}}}		
\mathnotation{\metric}{g}			
\mathnotation{\detmetric}{|\metric|}		
\mathnotation{\lyap}{\lambda}
\mathnotation{\nugr}{\Lambda}			
\mathnotation{\nugrpara}{\widetilde\nugr}	%
\mathnotation{\curvc}{\kappa}			
\mathnotation{\curvv}{\boldsymbol{\curvc}}	
\mathnotation{\curvm}{\curvc}			

\mathnotation{\fbend}{f}			
\mathnotation{\gbend}{h}			
\mathnotation{\aconst}{c}			
\mathnotation{\Cline}{{\mathcal{C}}}		
\mathnotation{\Len}{L}				
\mathnotation{\Carclen}{s}			
\mathnotation{\dCarclen}{ds}			
\mathnotation{\tanc}{\ell}			
\mathnotation{\tanuc}{{\hat\ell}}		
\mathnotation{\tanv}{{\boldsymbol{\tanc}}}	
\mathnotation{\tanuv}{{\hat{\boldsymbol{\tanc}}}}
\mathnotation{\vecc}{w}				
\mathnotation{\vecv}{{\mathbf{\vecc}}}		
\mathnotation{\upar}{s}				
\mathnotation{\Kstmap}{K}			

\newcommand{\twoD}{2D}
\newcommand{\threeD}{3D}

\begin{document}

\title{Stretching and Curvature of Material Lines in Chaotic Flows}
\author{Jean-Luc Thiffeault}
\date{\today}
\email{jeanluc@mailaps.org}
\affiliation{Department of Applied Physics and Applied Mathematics,
Columbia University, New York, NY 10027}
\altaffiliation[Present address: ]{Department of Mathematics, Imperial College
  London, SW7 2AZ, United Kingdom.}

\begin{abstract}
As a streak of dye is advected by a chaotic flow, it stretches and folds and
becomes indistinguishable from a one-dimensional idealized material line.  The
variation along a material line of the total stretching experienced by fluid
elements is examined, and it is found that it can be decomposed into an
overall time-dependent factor, constant along the line, and a smooth
time-independent deviation.  The stretching is related by a power law to the
curvature of the line near sharp bends.  This is confirmed numerically and
motivated by a simple model.  A conservation law for Lyapunov exponents
explains deviations from a power-law.
\end{abstract}

\maketitle

\section{Introduction}

The literature on the deformation of fluid elements in chaotic and turbulent
flows is immense.  Mainly the focus has been on the Lyapunov exponents, which
characterize the exponential rate of separation of neighboring trajectories,
and thus also represent the rate of stretching of fluid elements.  For
instance, the distribution of Lyapunov exponents has been used to characterize
the diffusive decay of variance of an advected
scalar~\cite{Pierrehumbert1991,Antonsen1991,Chertkov1995,Antonsen1996,%
Son1999,Balkovsky1999}, because in incompressible flows the stretching of
fluid elements has an associated contraction and thus serves to amplify
gradients, thus enhancing the efficiency of diffusion.

Much less attention has been paid the evolution of curvature in chaotic and
turbulent flows.  The curvature represents a higher-order deformation of fluid
elements than stretching.  To leading order, one can think of the stretching
as deforming an initially spherical fluid element into an ellipsoid; the
curvature then represents a bending of the axes of the ellipsoid.  Since
stretching tends to occur along a dominant direction, so the effect of
curvature is most easily observed along that dominant direction of stretching.
Mathematically, the stretching depends on first derivatives of the velocity
field (essentially the local strain), whereas curvature depends on its first
and second derivatives.

It has been observed that the magnitude of stretching and of curvature are
anticorrelated in a flow~\cite{Drummond1991}: wherever the flow experiences
large amounts of stretching the fluid elements (and hence material lines) tend
to be straight, and vice-versa.  Intuitively, one can see that this is a
result of the competing effects of curvature and stretching: the flow can pull
on a material line or bend it, but the effects are orthogonal---stretching is
the result of a pull along the material line, whereas curvature arises from
perpendicular deformations.  This provides a motivation for the study of
curvature in a flow, because being a purely geometrical quantity it can be
measured directly from pictures of material lines (which can easily be
obtained from experiments).  In this manner one can get a rough estimate of
regions of high and low stretching.  This is a useful guide when trying to
maximize the efficiency of mixing by varying the properties of the
flow~\cite{Hobbs1998b,Tang1999a}.

In the present paper we explore the form of the dependence between stretching
and curvature along material lines.  There is a~$-1/3$ power-law relation
between the magnitude of stretching and curvature along sharp bends of
material lines advected by a flow.  We present evidence for this based on
numerical experiments.  We then explain the~$-1/3$ power law using two models.
The first is based on a single sharp bend in a material line where only the
shape of the bend is taken into account; it has the advantage of being
straightforward and intuitive, but does not account for the range of features
observed.  The second model uses a foliation of bends, to correct the
deficiency of the first model of treating material lines as isolated objects
in the flow.  We make use of a local ``conservation law'' for Lyapunov
exponents that allows a more complete description of the relationship between
curvature and stretching for the foliation of bends.  In particular, the
conservation law predicts a dependence on the shape of neighboring material
lines near a particular bend.

\section{Stretching of a Material Line}
\label{sec:stretching}

We first discuss the kinematics of stretching of a material line advected by a
flow~\hbox{$\xv = \flow(\time,\time_0;\lav)$} in an~$\sdim$-dimensional space.
Here~$\xv$ is the position at time~$\time$ (the Eulerian coordinate) of a
fluid element that was originally at~$\lav$ at time~$\time_0$ (the Lagrangian
coordinate).  The flow~$\flow$ is typically the result of the integration of a
velocity field~$\velv(\xv,\time)$, but it could also be given by a map.

A vector~$\vecv_0$ is transformed to a vector~$\vecv$ at time~$\time$ by the
relation
\begin{equation}
	\vecc^i(\xv,\time) = {\Mc^i}_q\,\vecc_0^q(\lav,\time_0),\qquad
	{\Mc^i}_q = {[\Mt]^i}_q
	\ldef \frac{\pd\flow^i}{\pd\lac^q}(\time,\time_0;\lav),
\end{equation}
where repeated indices are summed; $\Mt$ is known as the tangent mapping of
the flow.  Because we are interested in the state of material lines in the
Eulerian ($\xv$) frame, we wish to hold~$\xv$ constant and integrate backward
in time to obtain~\hbox{$\lav = \flow^{-1}(\time,\time_0;\xv)$} (if~$\flow$ is
a map we iterate its inverse).  We thus rewrite the tangent mapping as a
function of~$\xv$,
\begin{equation}
	\Mt =
	\frac{\pd\flow}{\pd\lav}(\time,\time_0;\flow^{-1}
		(\time,\time_0;\xv))
	= \l\{\frac{\pd\flow^{-1}}{\pd\xv}(\time,\time_0;\xv)\r\}^{-1},
\end{equation}
which can easily be verified by differentiating the
identity~$\flow(\time,\time_0;\flow^{-1}(\time,\time_0;\xv))=\xv$ with respect
to~$\xv$.
The matrix~$\Mt$ is obtained directly from integrating
\begin{equation}
	\frac{\pd}{\pd\time_0}\Mt
	= -\Mt\cdot\transp{(\grad\velv)},\qquad
	\Mt(\time_0 = \time) = I,
\end{equation}
as~$\time_0\rightarrow-\infty$; here~$\grad\velv$ is evaluated
at~\hbox{$\lav = \flow^{-1}(\time,\time_0;\xv)$}.  In practice, there are more
accurate numerical methods available, based on matrix decomposition
techniques~\cite{Greene1987,Goldhirsch1987,Thiffeault2002}.

The stretching of vectors as they are advected by the flow is expressed in
the Eulerian frame by the left Cauchy--Green tensor~\cite{Ottino},
\begin{equation}
	\metric^{ij} \ldef \sum_{q=1}^\sdim\,{\Mc^i}_q{\Mc^j}_q\,.
\end{equation}
Being a symmetric positive-definite matrix, the Cauchy--Green tensor~$\metric$
admits~$\sdim$ real positive eigenvalues~$\nugr_\mu^2(\xv,\time,\time_0)$ with
corresponding orthonormal eigenvectors~$\ediruv_\mu(\xv,\time,\time_0)$.  We
assume that the eigenvalues are nondegenerate (at least for
large~$\time-\time_0$), and without loss of generality order them such
that~$\nugr_1 > \nugr_2 > \ldots > \nugr_\sdim$.  The~$\nugr_\mu$'s are called
the \emph{coefficients of expansion}, and their exponential growth rates
\begin{equation}
	\lyap_\mu \ldef \frac{1}{\time-\time_0}\,\log\nugr_\mu
\end{equation}
are known as the finite-time Lyapunov exponents.  The theorem of
Oseledec~\cite{Oseledec1968} asserts that for ergodic measure-preserving
systems the limit as~$\time\rightarrow\infty$ (or in our
case~$\time_0\rightarrow-\infty$) exists for almost all initial conditions (or
in our case the final condition~$\xv$ at~$\time$, held fixed).  The positivity
of the largest Lyapunov exponent is the usual criterion for the presence of
chaos.  Because the eigenvalue of largest stretching plays an important role
in our development, we often write~$\nugr_\udir$ and~$\udiruv$ for~$\nugr_1$
and~$\ediruv_1$, where the letter~$\udir$ stands for ``unstable''.  Note that
we shall not assume that~$\lyap_\udir$ has converged, only that it is positive
``most of the time,'' in the sense that~$\nugr_1$ is exponentially large for
large~$\time-\time_0$.  Thus our treatment is valid even for aperiodic flows,
where the Lyapunov exponents are not guaranteed to converge.

The eigenvector~$\udiruv$ is the direction along which a fluid element has on
average experienced the most stretching throughout its history; this direction
converges exponentially to an asymptotic direction~$\udiruvinf(\xv,\time)$
as~$\time_0\rightarrow-\infty$~\cite{Greene1987,Goldhirsch1987,Thiffeault2002}.
We can integrate this vector field,
\begin{equation}
	\frac{\pd\xv_\udir}{\pd\upar} =
	\udiruvinf(\xv_\udir(\upar,\time),\time),\qquad
	\udiruvinf(\xv_\udir(0,\time),\time) = \udiruvinf(\barxv ,\time),
	\label{eq:umanode}
\end{equation}
to yield a curve~$\xv_\udir(\upar,\time)$ through~$\barxv$.  This (infinite)
curve is known as the \emph{global unstable manifold} through~$\barxv$, and
is parametrized by the arc length~$\upar$ along the curve from~$\barxv$.  It
corresponds to the unstable manifold of a hyperbolic orbit and converges to
that orbit as~$\time_0\rightarrow-\infty$.  In a given chaotic region, all
global unstable manifolds are equivalent to each other~\cite{Giona1998}; this
equivalence class of unstable manifolds is called the \emph{unstable
foliation}.  We will think of the representative point~$\barxv$ as labeling a
particular unstable manifold.

Now consider a material line~$\Cline$ at time~$\time$, and
let~$\tanuv(\xv,\time)$ be the unit tangent to the line~$\Cline$ at~$\xv$.  It
is well-known that a material line advected by a chaotic flow aligns with the
unstable foliation of the flow, a phenomenon sometimes referred to as
\emph{asymptotic directionality}~\cite{Giona1998}.  Intuitively, it is clear
that if fluid elements are being stretched along a preferred direction then
they appear to align along that direction.  This means that the
tangent~$\tanuv$ aligns with the most unstable eigenvector~$\udiruvinf$ of the
flow as~$\time_0\rightarrow-\infty$.  The components of~$\tanuv$ initially
orthogonal to~$\udiruvinf$ decay in proportion to their slower stretching
rates as compared to~$\nugr_\udir$.  This can be expressed as
\begin{equation}
	\tanuv(\xv,\time,\time_0) = \sum_{\nu=1}^\sdim
	\frac{\nugr_\nu}{\nugr_\udir}\,
	\sigma_{\nu}(\lav,\time,\time_0)\,
	\ediruvinf_\nu(\xv,\time)
	\label{eq:tanedir}
\end{equation}
where the~$\sigma_\nu$ are nonexponential functions, that is, they may grow or
decay algebraically (or be identically zero) but may not do so exponentially;
they are given by the initial condition of~$\tanuv(\xv,\time_0,\time_0)$.  The
normalization condition of~$\tanuv$ implies
\begin{equation}
	\sigma_1 = \pm\Bigl(1 - \sum_{\nu>1}(\nugr_\nu/\nugr_\udir)^2\,
		\sigma_{\nu}^2\Bigr)^{1/2},
\end{equation}
where the sign ambiguity occurs because the direction of the alignment is a
matter of convention.  Thus~$|\sigma_1|\rightarrow 1$
as~$\time_0\rightarrow-\infty$, whereas the other~$\sigma_\nu$ decay
exponentially as~$\nugr_\nu/\nugr_\udir$, a consequence of the ordering and
nondegeneracy assumptions for the~$\nugr_\nu$'s.

Let~$\dCarclen$ denote the infinitesimal element of arc length at some
point on the line.  To measure the relative growth of this element, we need to
know its length~$\dCarclen_0$ at an earlier time~\hbox{$\time_0 < \time$},
\begin{equation}
	\dCarclen_0
	= \l({\metric^{-1}_{ij}\,\tanuc^i\,\tanuc^j}\r)^{1/2}\dCarclen.
	\label{eq:dst0}
\end{equation}
The square-root in~\eqref{eq:dst0} is the
Jacobian~$|\pd\Carclen_0/\pd\Carclen|$, since the arc length changes by the
same proportion as the length of the tangent.  Inserting the
tangent~\eqref{eq:tanedir} in the expression~\eqref{eq:dst0} for the length
element, we find
\begin{equation}
	\dCarclen_0 =
	\biggl(\sum_\nu \frac{\nugr_\nu^2}{\nugr_\udir^2}\,\sigma_\nu^2\,
		\nugr_\nu^{-2}\biggr)^{1/2}\dCarclen
	= \nugr_\udir^{-1}\biggl(\sum_\nu \sigma_\nu^2\,\biggr)^{1/2}
		\dCarclen,
	\label{eq:dst0u}
\end{equation}
where we have used~\hbox{$\metric_{ij}^{-1}\,(\ediruinf_\nu)^j =
\nugr_\nu^{-2}\,(\ediruinf_\nu)^i$} from the definition of the eigenvectors.
Because of the chaotic nature of the flow, $\nugr_\udir =
\exp(\lyap_\udir\,\time)$ grows exponentially in time,
with~$\lyap_\udir(\xv,\time)$ the largest finite-time Lyapunov exponent.  In
general, then, the ratio~$\dCarclen/\dCarclen_0$ grows exponentially
with~$\time-\time_0$, reflecting the growth of material lines as they are
advected by the flow.

The local stretching of the line,~$\dCarclen/\dCarclen_0$, varies both because
of the inherent nonuniformity in infinitesimal stretching, as described
by~$\nugr_\udir$, but also because the initial material line at~$\time_0$ is
typically aligned differently with respect to the global unstable manifold at
different points, as described by the~$\sigma_\nu$'s.  However, the first
effect is exponential in time, whereas the second is only algebraic.  We can
further differentiate these temporal behaviors, as we now proceed to show.

If~$\time-\time_0$ is moderately large then the material line has aligned with
the global unstable manifold, that is, it is tangent to~$\udiruvinf$ at every
point.  The relative change in~$\nugr_\udir^{-1}$ along~$\tanuv$ satisfies
\begin{equation}
	\tanuv\cdot\grad\log\nugr_\udir^{-1}
		= \tanuv\cdot\grad\log\nugrpara_\udir^{-1}
	+ \Order{\nugr_\udir^{-1},\nugr_2/\nugr_\udir},
	\qquad\text{for $|\time-\time_0|\gg 1$},
	\label{eq:nugrudirvar}
\end{equation}
where~$\nugrpara_\udir$ depends on~$\xv$ and~$\time$ but not on the initial
time~$\time_0$.  (We define~$\nugrpara_\udir$ more precisely below
in~\eqref{eq:nugrdecomp}.)  This asymptotic form is a consequence of the
results on derivatives of finite-time Lyapunov of Ref.~\cite{Thiffeault2002}.
It is also straightforward to show that the relative change in~$\sigma_\nu$ is
\begin{equation}
	\tanuv\cdot\grad\log\sigma_\nu
		= \Order{\nugr_\udir^{-1}},
	\qquad\text{for $|\time-\time_0|\gg 1$},
	\label{eq:tanuvvar}
\end{equation}
consistent with the~$\sigma_\nu$ being defined in Lagrangian coordinates, so
the stretching along~$\udiruvinf$ in Eulerian coordinates smooths out their
functional dependence.

Integrating equation~\eqref{eq:nugrudirvar} allows us to write
\begin{equation}
	\nugr_\udir(\xv,\time,\time_0) \simeq
		\nugr_\udir(\barxv,\time,\time_0)\,
		\nugrpara_\udir(\xv,\time),\qquad
	\text{for $|\time-\time_0|\gg 1$},
	\label{eq:nugrdecomp}
\end{equation}
where~$\nugr_\udir(\barxv,\time,\time_0)$ is the coefficient of expansion
evaluated at an arbitrary reference point~$\barxv$ on the
manifold~$\xv_\udir(\upar,\time)$, defined by Eq.~\eqref{eq:umanode}.  This is
true as long as we are considering a line segment~$\Cline(\time)$ that is
shorter than~$\nugr_\udir$, otherwise the higher-order terms
in~\eqref{eq:nugrudirvar} cannot be neglected.  The asymptotic exponential
dependence of~$\nugr_\udir$ on~$\time-\time_0$ is entirely contained
in~$\nugr_\udir(\barxv,\time,\time_0)$, because~$\nugrpara$ does not depend
on~$\time_0$.

Similarly, Eq.~\eqref{eq:tanuvvar} implies that the~$\sigma_\nu$ are constant
in space along~$\Cline(\time)$ as long as it is shorter than~$\nugr_\udir$.
We can then absorb the constant~$(\sum\sigma_\nu^2)^{1/2}$
into~$\nugr_\udir(\barxv,\time,\time_0)$ and rewrite~\eqref{eq:dst0u} as
\begin{equation}
	\dCarclen/\dCarclen_0
	= \nugr_\udir(\barxv,\time,\time_0)\,
		\nugrpara_\udir(\xv,\time),
	\label{eq:dst0uu}
\end{equation}
where~$\nugr_\udir(\barxv,\time,\time_0)$ is constant in space along the
line~$\Cline$.

The main result of this section is thus given by~\eqref{eq:dst0uu}.  In the
remainder of this paper we shall examine line segments at fixed~$\xv$
and~$\time$, letting~$\time_0$ recede to~$-\infty$.  Since the segments we
consider are much shorter than~$\nugr_\udir$, we find that the local
stretching (coefficient of expansion) of a material line can be characterized
by the deviation~$\nugrpara_\udir(\xv,\time)$ from the overall stretching of
the line segment,~$\nugr_\udir(\barxv,\time,\time_0)$.  The somewhat
surprising results are that (i) the deviation~$\nugrpara_\udir(\xv,\time)$ is
independent of~$\time_0$, and (ii) the overall stretching of the line is
well-defined (\ie, it is constant along the line).  The rate of overall
stretching of the line of course converges to the topological entropy.

We note in closing this section that there is an ambiguity in the
decomposition~\eqref{eq:nugrdecomp}:~$\nugr_\udir(\barxv,\time,\time_0)$
and~$\nugrpara_\udir(\xv,\time)$ are only defined up to a multiplicative
function of~$\barxv$ and~$\time$.  We resolve this by
integrating~\eqref{eq:dst0uu} over~$\Cline$, to find
\begin{equation}
	\Len(\time)/\Len(\time_0)
	= \nugr_\udir(\barxv,\time,\time_0)\,
		\l\langle\nugrpara_\udir^{-1}\r\rangle^{-1}_{\Cline(\time)},
	\label{eq:LtLt0}
\end{equation}
where~$\Len(\time)$ is the length of~$\Cline(\time)$
and~$\langle\cdot\rangle_{\Cline(\time)}$ is an average over the
line~$\Cline(\time)$.  Clearly a proper definition of the overall
growth~$\nugr_\udir(\barxv,\time,\time_0)$ of the material line should be
that it is equal to the growth of the length~$\Len(\time)/\Len(\time_0)$.  An
appropriate choice for the definition of~$\nugr_\udir(\barxv,\time,\time_0)$
and~$\nugrpara_\udir(\xv,\time)$ is thus to scale them such that
\begin{equation}
	\l\langle\nugrpara_\udir^{-1}\r\rangle_{\Cline(\time)} = 1,
	\label{eq:nugrparanorm}
\end{equation}
which we assume to hold.

\section{Numerical Results}

In this section we present the result of numerical calculations comparing the
stretching and the curvature along material lines.  Because we are interested
in material lines that have evolved for a long time in the flow (long compared
to the inverse Lyapunov exponent), it is sufficient to compute the shape of
unstable manifolds by integrating~\eqref{eq:umanode} starting at sample
points~$\xv$ at time~$\time$ in the fluid domain.

\begin{figure*}
\centering
\psfrag{x}{$x$}
\psfrag{y}{$y$}
\subfigure[]{
	\includegraphics[width=.45\textwidth]{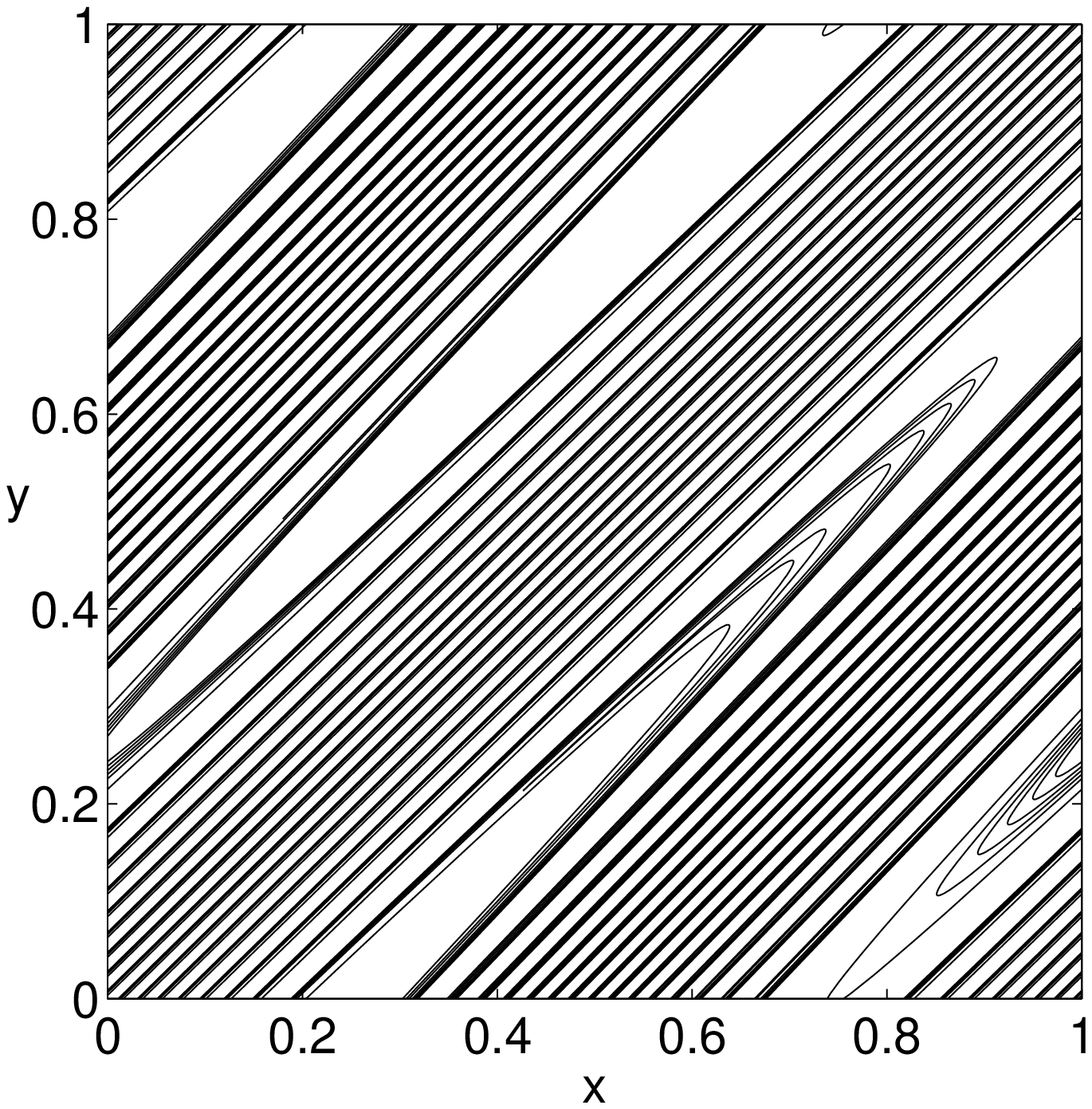}
	\label{fig:uman_stmap}
}\goodgap
\subfigure[]{
	\includegraphics[width=.45\textwidth]{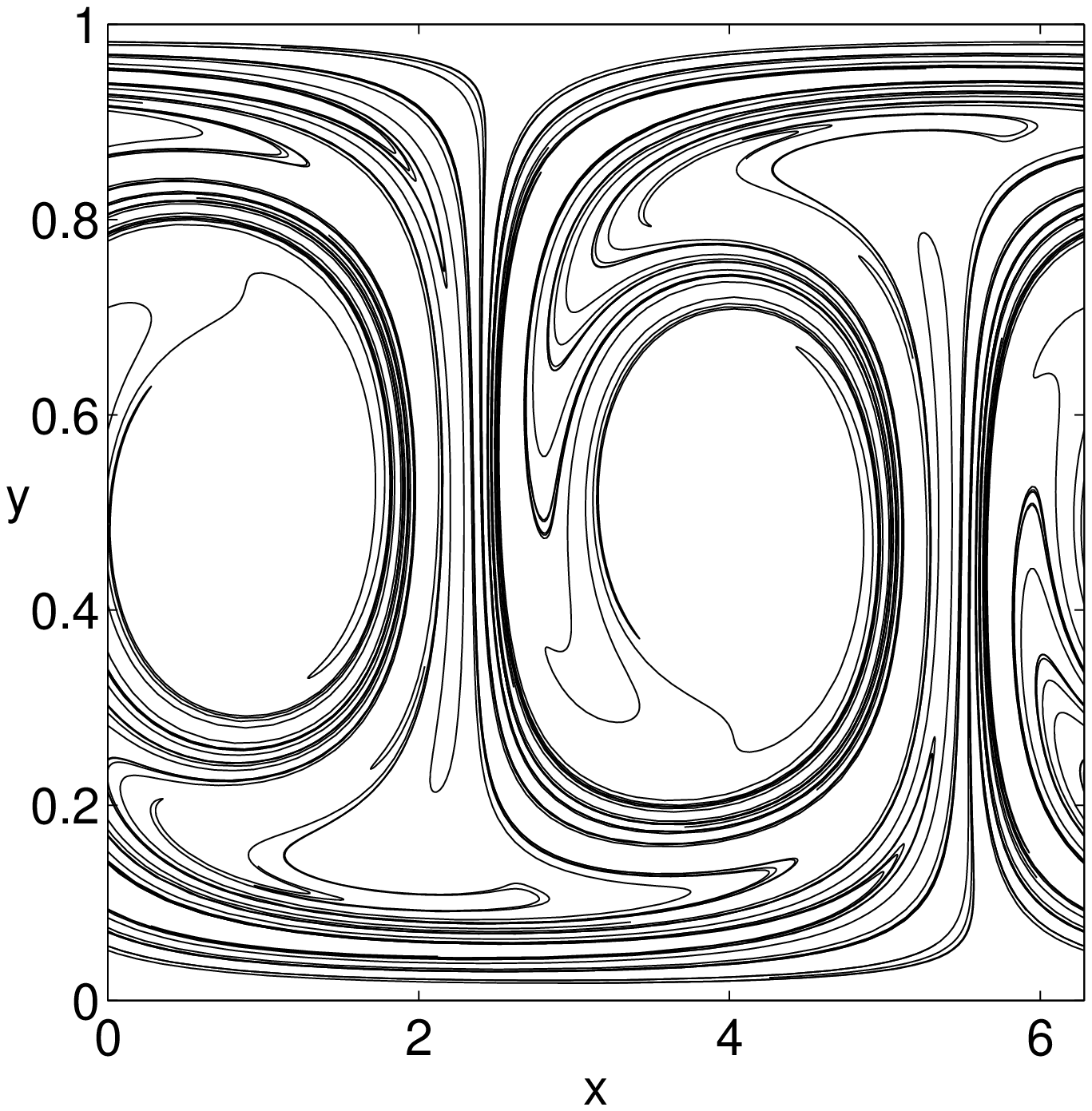}
	\label{fig:uman_crollsns}
}
\caption{Material lines advected by (a) the standard map with~$\Kstmap=50$ and
(b) the cellular flow with~$A=\epsilon=\omega=k=1$.}
\label{fig:uman}
\end{figure*}

Two prototypical systems will be used.  The first is the ubiquitous standard
map,
\begin{equation}
	x_{n+1} = x_n + y_{n+1}\,,\qquad
	y_{n+1} = y_n + (\Kstmap/2\pi)\,\sin(2\pi x_n),
	\label{eq:standardmap}
\end{equation}
where~$(x,y) \in [0,1]\times[0,1]$.  Maps of course have the advantage that
they are computationally efficient, allowing rapid verification of results.
Another desirable feature of~\eqref{eq:standardmap} is that for
large~$\Kstmap$ the structure of material lines advected by the map is fairly
simple, as evidenced in Fig.~\ref{fig:uman_stmap} which has~$\Kstmap=50$.
The diagonal bias in the figure is due to the sweeping of the sine wave on
both side of~\eqref{eq:standardmap}, which at such large~$\Kstmap$ traverses
the domain several times for each iteration, breaking (almost) all islands.
The map is area-preserving, so that the determinant of~$\metric$
is~$\detmetric=1$.

The second system we shall use is the cellular flow of Solomon and
Gollub~\cite{Solomon1988}, with velocity field
\begin{equation}
	\velv
	\ldef \l(-\frac{\pd\psi}{\pd y},\frac{\pd\psi}{\pd x}\r),\qquad
	\psi(\xv,\time) \ldef A\,k^{-1} (\sin k x
		+ \epsilon \cos\omega \time\,\cos k x) \sin^2 \pi y,
	\label{eq:crollsns}
\end{equation}
to model an array of oscillating convection rolls, periodic in~$x$ and with
rigid walls at~$y=0$ and~$y=1$.  The~$\sin^2 \pi y$ dependence is chosen to
satisfy the rigid boundary conditions at the walls.  The velocity field is
incompressible ($\div\velv=0$), so that~$\detmetric=1$.  When $\omega=0$, the
flow is steady, and the trajectories of fluid elements are nonchaotic.  This
is true in general of any two-dimensional steady flow~\cite{Eckmann1985}.  A
snapshot of typical material lines advected by~\eqref{eq:crollsns} are shown
in Fig.~\ref{fig:uman_crollsns}.  Note the two large regular islands in the
center.

Our interest lies in the variations of stretching and curvature along material
lines.  The local stretching is given by the coefficient of
expansion~$\nugr_\udir$, but since we let the material lines evolve for some
time (such that~$\nugr_\udir\gg 1$) it is sufficient to consider the
$\time_0$-independent deviation from mean stretching,~$\nugrpara_\udir$,
defined in Section~\ref{sec:stretching}.  Thus, for convenience we will refer
to~$\nugrpara_\udir$ as ``the stretching'' even though it needs to be
multiplied by an overall factor~$\nugr(\barxv,\time,\time_0)$ to represent
the total stretching a line has experienced.

\begin{figure*}
\psfrag{th}{$\Carclen$}
\psfrag{log k}{\raisebox{.7ex}{$\log\curvm$}}
\psfrag{log L}{\raisebox{.7ex}{$\!\!\!\!\log\nugrpara_\udir$}}
\subfigure[]{
	\includegraphics[height=.36\textwidth]
		{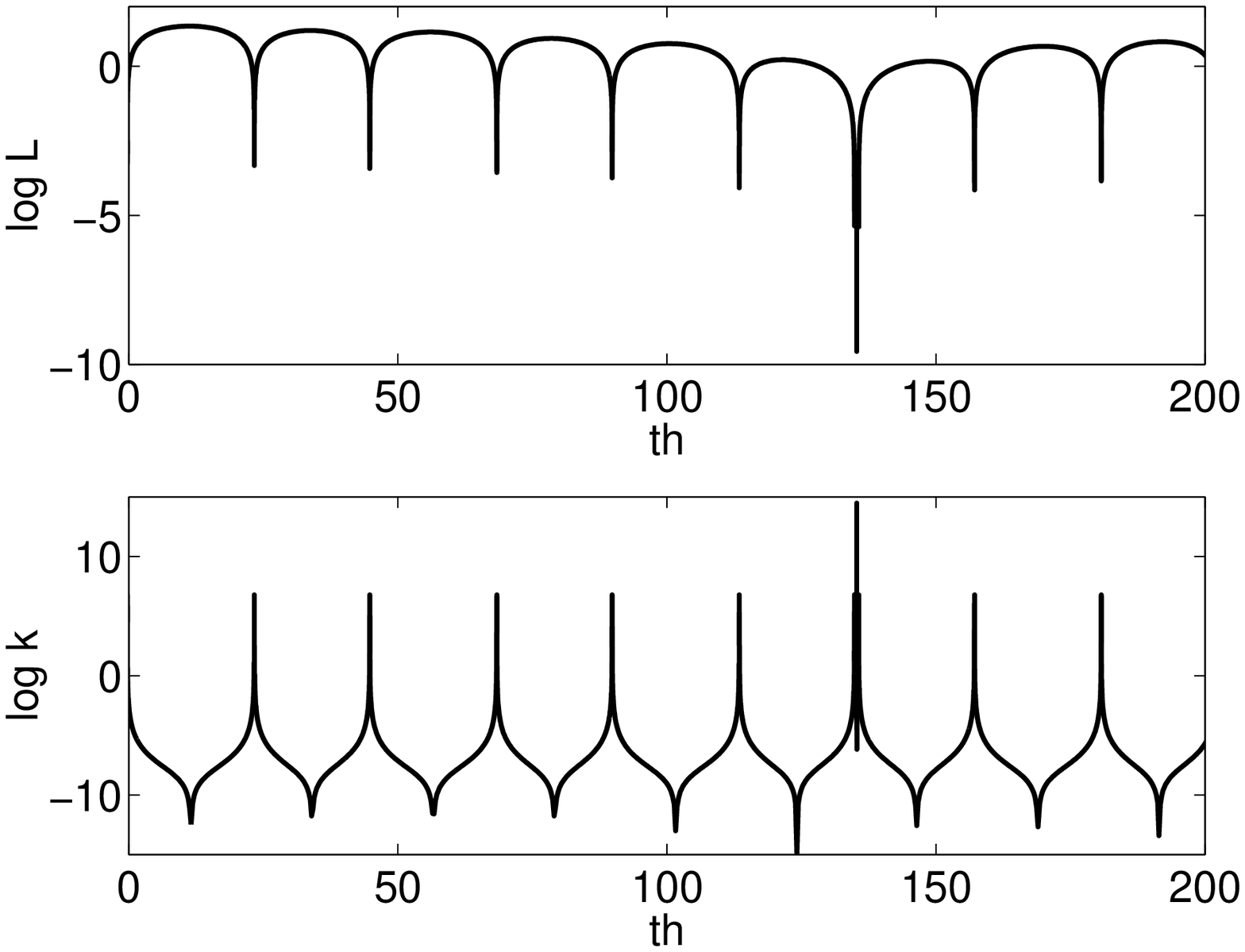}
	\label{fig:Lamt_curv_vs_tauh_stmap}
}\goodgap
\subfigure[]{
	\includegraphics[height=.36\textwidth]
		{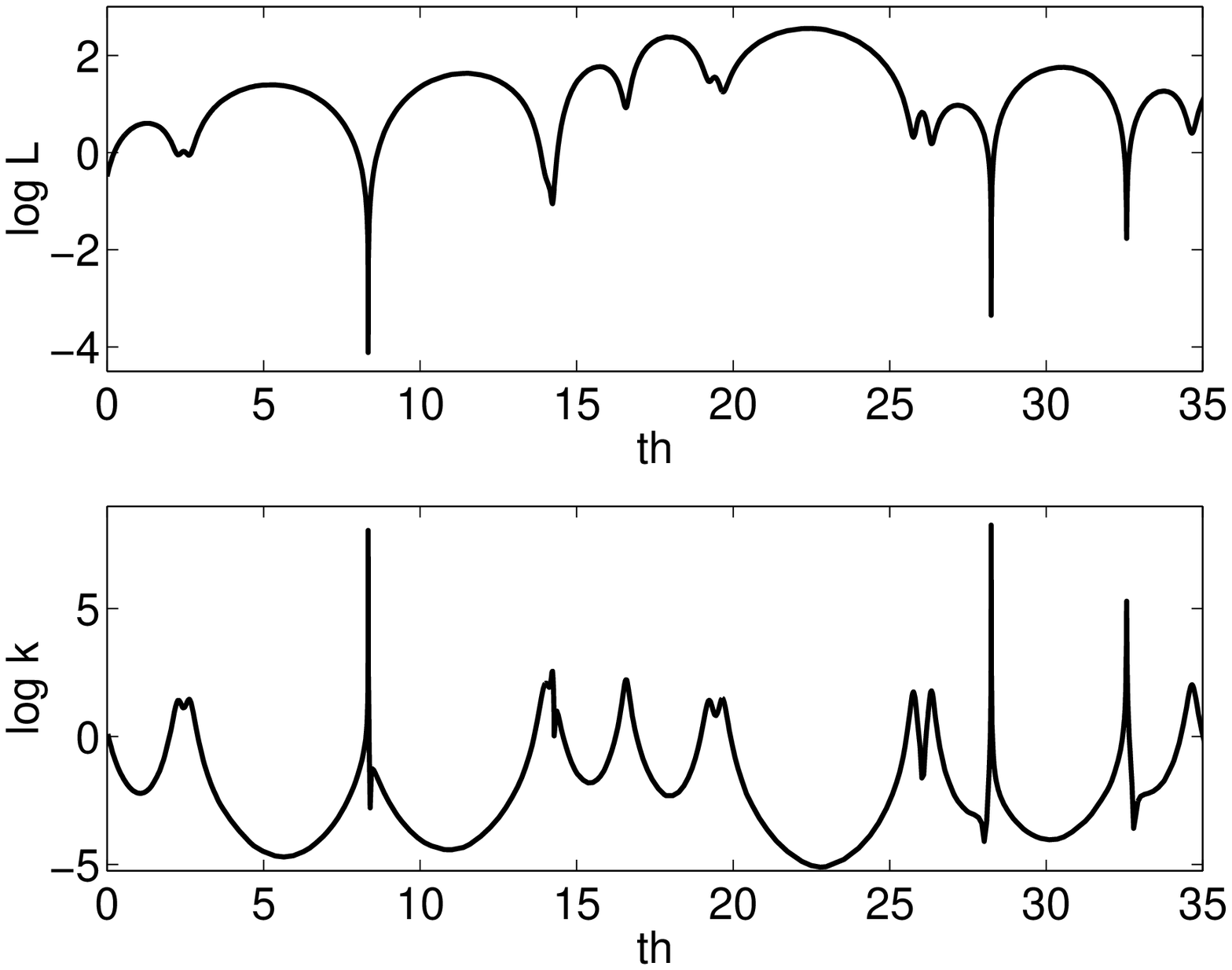}
	\label{fig:Lamt_curv_vs_tauh_crollsns}
}
\caption{The stretching,~$\nugrpara_\udir$, and the magnitude of the
curvature,~$\curvm$, plotted against the arc length~$\Carclen$ along a
material line for (a) the standard map; and (b) the cellular flow.  The
parameter values are as in Fig.~\ref{fig:uman}.  Note the close
anticorrelation between stretching and curvature.}
\label{fig:Lamt_curv_vs_tauh}
\end{figure*}

Figure~\ref{fig:Lamt_curv_vs_tauh}
is a plot of the stretching and curvature as they vary along a typical
material line advected by our two prototypical systems.  The first thing to
note is that the stretching~$\nugrpara_\udir$ is smooth along the material
line (the sharp dips are well-resolved).  It is known that the finite-time
Lyapunov exponents are not continuous functions of the spatial variables, but
Fig.~\ref{fig:Lamt_curv_vs_tauh} underscores that the rough variation of the
exponents (and hence the stretching) occurs perpendicular to the unstable
direction.  In the Lagrangian frame, holding~$\lav$ fixed instead of~$\xv$, it
is the variation along the contracting direction that is
smooth~\cite{Tang1996,Thiffeault2001}.  (The variation of the stretching along
the characteristic directions is discussed in Ref.~\cite{Thiffeault2002}.)

A second observation is that the variations in the stretching are very large,
encompassing about four orders of magnitude or more in typical cases.  This
says that not all points in the flow are equally propitious to stretching, and
that the variations can be sizeable even within a single chaotic region.
Because the coefficient of expansion denotes a fluid elements's history of
stretching, the large variations imply that some trajectories avoid regions of
significant stretching for a long time.

The third striking feature of Fig.~\ref{fig:Lamt_curv_vs_tauh} is the close
anticorrelation between stretching and curvature---high curvature regions are
invariably associated with relatively low stretching.  This phenomenon was
first observed by Drummond and M\"unch~\cite{Drummond1991} in the context of a
model turbulent flow, based on earlier work of Pope~\etal\ on the curvature of
fluid elements~\cite{Pope1988,Pope1989}.  This and later work on turbulence
and random flows~\cite{Ishihara1992,Drummond1993,Liu1996,Schekochihin2002} and
deterministic flows~\cite{Liu1996,Hobbs1997b,Hobbs1998,Cerbelli2000} focused
on the probability distribution of curvature and on comparing stretching and
curvature at a point.  Here we investigate how stretching and curvature vary
along material lines, with the intention of gaining a better understanding of
their relationship.

\begin{figure*}
\psfrag{log k}{\raisebox{-1ex}{$\log\curvm$}}
\psfrag{log L}{\raisebox{1ex}{$\!\!\!\!\log\nugrpara_\udir$}}
\psfrag{k3a}{$\!\!\!\!\!\!\!\!\!\!\curvm^{-1/3}$}
\psfrag{k3b}{$\curvm^{-1/3}$}
\psfrag{k3}{$\!\!\!\!\!\!\!\!\!\!\curvm^{-1/3}$}
\subfigure[]{
	\includegraphics[height=.37\textwidth]{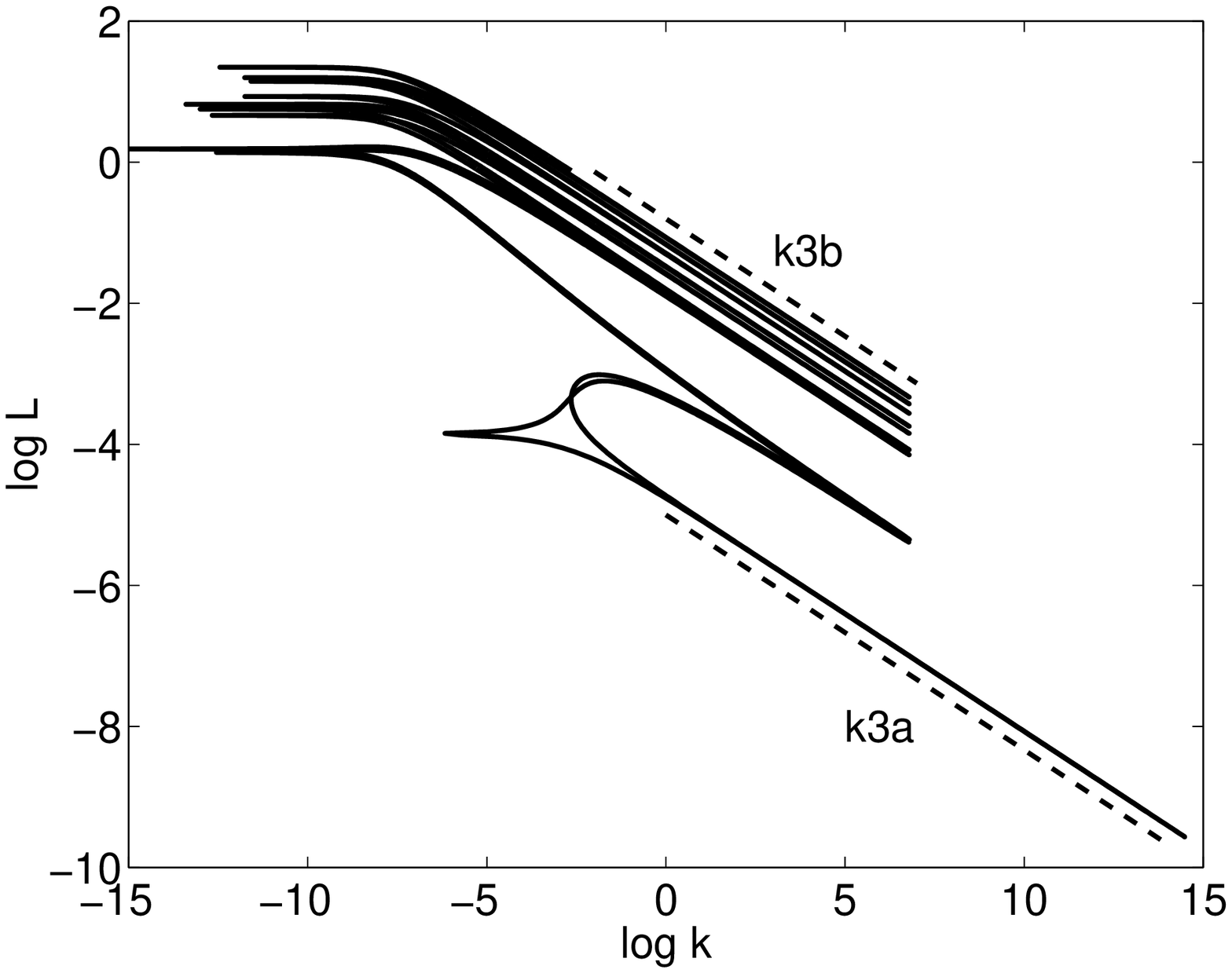}
	\label{fig:Lamt_vs_curv_stmap}
}\goodgap
\subfigure[]{
	\includegraphics[height=.37\textwidth]{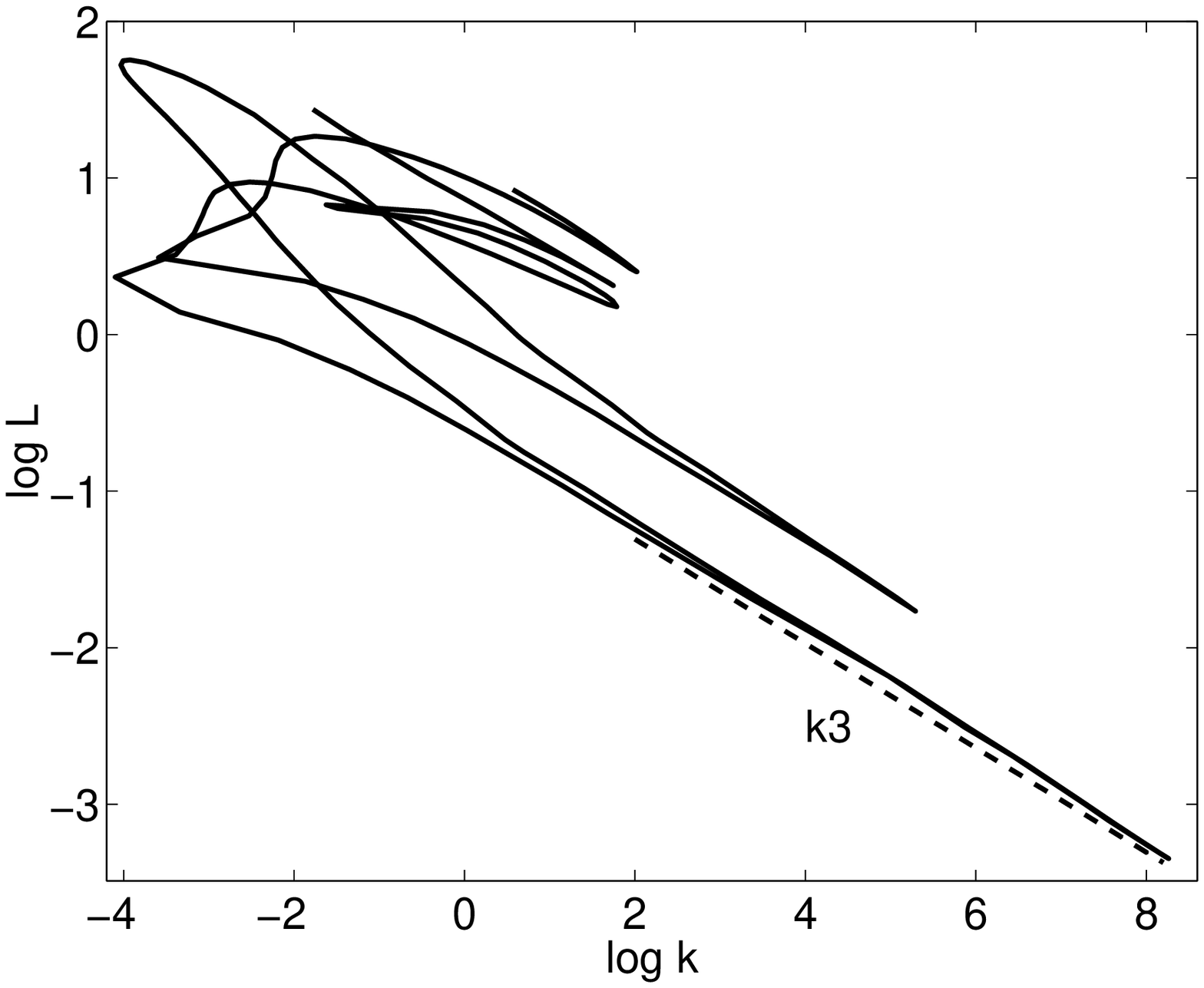}
	\label{fig:Lamt_vs_curv_crollsns}
}
\caption{Parametric plot of the stretching,~$\nugrpara_\udir$, against the
magnitude of the curvature,~$\curvm$, along a material line for (a) the
standard map; and (b) the cellular flow.  The parameter values are as in
Fig.~\ref{fig:uman}.  For sharp bends there is a power-law relationship
between the curvature and the stretching.  From (a), it is clear that for the
same maximum curvature, bends can have very different peak stretching factors.
}
\label{fig:Lamt_vs_curv}
\end{figure*}

With this in mind it is natural to plot the stretching and curvature of
Fig.~\ref{fig:Lamt_curv_vs_tauh} as a parametric (or phase) plot, that is, we
plot them against each other while increasing the arc length along the line.
The result is shown in Fig.~\ref{fig:Lamt_vs_curv}.
A power-law relation between stretching and curvature around sharp bends
(large curvature) is immediately apparent, with exponent~$-1/3$.  This
power-law behavior is especially well-realized for the standard map
(Fig.~\ref{fig:Lamt_vs_curv_stmap}).  Fig.~\ref{fig:Lamt_vs_curv_stmap} also
shows that for the same maximum curvature, bends can have very different peak
stretchings, so even though there is a power-law relationship between
curvature and stretching, the peak stretching is not directly related to the
peak curvature.

In a related context involving the stable manifold in the Lagrangian frame,
the exponent was measured as~$-0.331$ by Tang and Boozer~\cite{Tang1996}.  In
the following two sections we use two models of increasing refinement to
justify an exponent of exactly~$-1/3$ and illuminate the cause of the
power-law relation.

\section{Analysis for a Single Bend}
\label{sec:simplebend}

We propose a simple physical picture in~\twoD\ to reproduce
the~$\nugrpara_\udir\sim\curvm^{-1/3}$ law around sharp bends.  We refer to
this picture as the ``simple bend model''.  We examine the model in detail,
and point out its advantages and failures.  Its advantages are simplicity and
a physically intuitive formulation.  Its failures are that it refers to a
specific initial configuration of the material line, and that it does not
account for the deviations from the~$-1/3$ law observed in
Fig.~\ref{fig:Lamt_vs_curv}; these shortcomings will be addressed by a more
refined model in Section~\ref{sec:conslaws}.

\begin{figure}
\centering
\subfigure[]{
	\psfrag{s}{\raisebox{-.25em}{\!\!\! $\Carclen$}}
	\psfrag{s=0}{{\!\!\! $\Carclen=0$}}
	\psfrag{0}{{$0$}}
	\psfrag{L}{{$\Len$}}
	\psfrag{bL}{$\Len'$}
	\psfrag{(a)}{}
	\psfrag{(b)}{}
	\psfrag{Fold}{{\!\! Fold}}
	\includegraphics[height=2.75in]{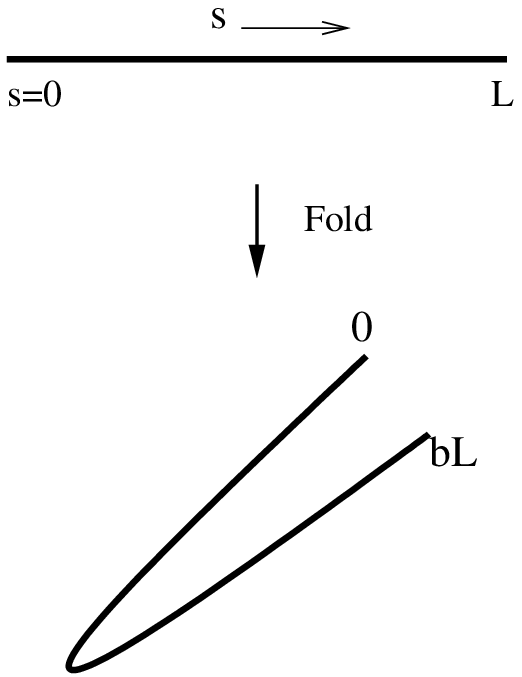}
	\label{fig:foldmech}
}\goodgap
\subfigure[]{
	\psfrag{D}{$\!\delta$}
	\psfrag{x}{$\!x$}
	\psfrag{y}{$\!\!\!\!y$}
	\includegraphics[height=2.75in]{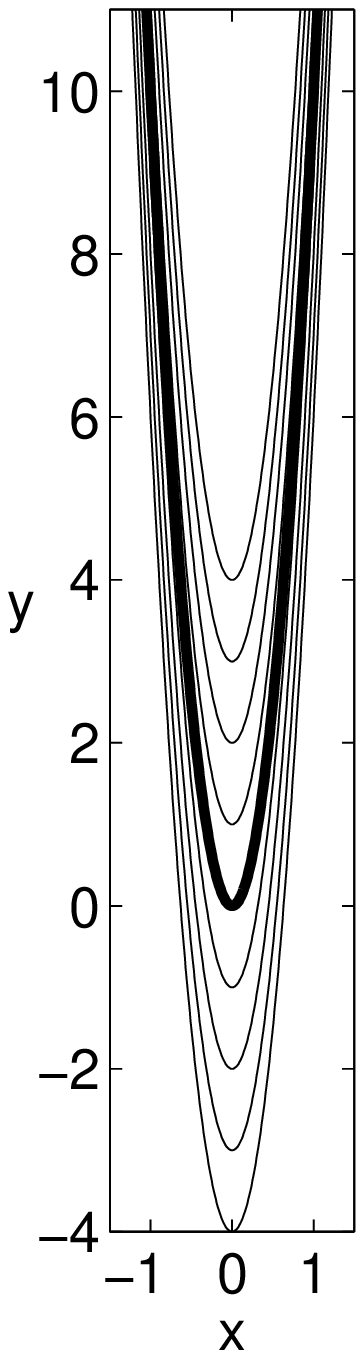}
	\label{fig:bendfoliation}
}
\caption{(a) Schematic representation of a material line folded by a flow.
(b) A foliation of bends.}
\label{fig:fold}
\end{figure}

Consider an initially straight material line of length~$\Len$, as illustrated
at the top of Fig.~\ref{fig:foldmech}, parametrized
by~\hbox{$(x(\Carclen),y(\Carclen)) = (\Carclen,0)$} with~\hbox{$\Carclen \in
[0,\Len]$}.
The material line will be stretched and folded by the flow, as in
Fig.~\ref{fig:foldmech}.  We parametrize the shape of the fold by
\begin{equation}
  \xv(s) = (x(\Carclen)\,,\,y(\Carclen)) =
  (\beta\Carclen\,,\,\fbend(\Carclen)),\qquad \Carclen \in [0,\Len],
  \label{eq:bendparam}
\end{equation}
where~$\fbend$ determines the shape of the bend, and~$\beta$ its horizontal
extent.  From this we can directly compute the curvature,
\begin{equation}
  \curvm = \frac{\l\lVert\xv'(\Carclen)\times\xv''(\Carclen)\r\rVert}
	 {\l\lVert\xv'(\Carclen)\r\rVert^3}
	 = \beta\,\l\lvert\fbend''(\Carclen)\r\rvert\,
	 \l\lVert\xv'(\Carclen)\r\rVert^{-3}.
  \label{eq:13law}
\end{equation}
But~$\l\lVert\xv'(\Carclen)\r\rVert$ is a measure of~$\nugrpara_\udir$, the
relative local stretching of the material line~\cite{Cerbelli2000}.  Hence, if
the variation in~$\l\lvert\fbend''(\Carclen)\r\rvert$ can be neglected,
Eq.~\eqref{eq:13law} gives the desired power-law relationship between
curvature and stretching.  That~$\l\lvert\fbend''(\Carclen)\r\rvert$ can be
neglected is a consequence of the sharpness of the bend: a Taylor series
expansion of~$\fbend$ (in a rotated and translated coordinate system)
\begin{equation}
	\fbend(\Carclen) = \tfrac{1}{2}\curvm_0\,\Carclen^2
		+ \mathrm{O}(\Carclen^3)
	\label{eq:bendexp}
\end{equation}
will be dominated by the quadratic term near the tip of the bend: the cubic
and other odd terms cannot dominate, otherwise the profile would not resemble
a sharp bend.  While it is possible that quartic or higher-order terms are
important in characterizing the shape of a given bend, we assume that there is
nothing special about this particular bend so that the quadratic term
dominates (in principle many shapes of bends will occur in a chaotic flow).
This will always be true if we are near enough the tip.  The
coefficient~$\curvm_0$ is large and gives the curvature at the tip of the
bend.

It is encouraging that such a straightforward physical picture reproduces the
salient features of stretching and folding in a chaotic flow.  But there are
problems with the simple bend model.  First, it requires a consideration of
the initial shape of the material line.  In a chaotic flow, the~$-1/3$ law is
observed for arbitrary initial conditions.  The second problem is that
deviations from the strict~$-1/3$ law are not of the form predicted by the
model.  In Fig.~\ref{fig:Lamt_vs_curv}, some of the bends appear ``fat'', that
is, the maxima of stretching and curvature do not exactly coincide.  The only
deviations allowed in the simple bend model arise from a non-quadratic bend
profile.  But the bends that exhibit deviations from the law in
Fig.~\ref{fig:Lamt_vs_curv} have a profile that is essentially quadratic near
the tip.

We resolve this discrepancy in Section~\ref{sec:conslaws} by allowing the bend
to be surrounded by a continuum (a \emph{foliation}) of other bends and using
a conservation law for Lyapunov exponents.

\section{A Conservation Law for Lyapunov Exponents}
\label{sec:conslaws}

The failures of the simple bend model in Section~\ref{sec:simplebend} lies in
the material lines not being isolated objects in the fluid: they are
surrounded by a continuum of other material lines.
Figure~\ref{fig:uman_stmap} shows a material line that has been advected by
the standard map, for a large value of the control parameter~$\Kstmap$ (by
``advection'' in the context of a map we simply mean iterating an initial
distribution).  At these highly chaotic values of~$\Kstmap$, the material line
exhibits a surprising degree of regularity.  Indeed, the visible folds (there
are also tiny bends too small to see) resemble a nested set of the simple bend
we discussed in Section~\ref{sec:simplebend}.  This suggests extending the
model to treat a \emph{foliation} of curves, as shown in
Fig.~\ref{fig:bendfoliation}, where the thick curve is the bend under
consideration (the nominal bend) and the others are neighboring bends (not
necessarily identical to the nominal bend).

The tangent~$\tanv(x,y)$ near the thick bend~$y=\fbend(x)$ of
Fig.~\ref{fig:bendfoliation} can be written
\begin{equation}
	\tanv(x,y) =
	\l(1\,,\,\fbend'(x) + \gbend(x,y-\fbend(x))\r),
	\qquad
	\gbend(x,0) = 0,
	\label{eq:fbendtangent}
\end{equation}
with the unit tangent~$\tanuv = \tanv/\lVert\tanv\rVert$.  Thus, when
evaluated on the curve~$y=\fbend(x)$, we have~\hbox{$\tanv =
(1\,,\,\fbend'(x))$} and we recover the tangent to the curve.  The
function~$\gbend$ represents changes in the tangent to the bends as we move
off our nominal bend~$y=\fbend(x)$.  Both~$\fbend$ and~$\gbend$ are left
unspecified for now.

To analyze the variation of stretching on the nominal bend, it is sufficient
to know how~$\nugr_\udir$ varies along~$\udiruvinf$, the characteristic
direction of stretching of fluid elements.  This is because material lines can
be assumed to be aligned with the unstable foliation, as described in
Section~\ref{sec:stretching}.  We will obtain this variation from the formula
\begin{equation}
	\detmetric^{1/2}\div\l(\detmetric^{-1/2}\,\udiruv\r)
	+ \udiruv\cdot\grad\log\nugr_\udir
	\sim \max(\nugr_\udir^{-1},\nugr_2/\nugr_\udir)
	\longrightarrow 0,
	\label{eq:constraint}
\end{equation}
which holds for for large~$\time-\time_0$, when we can replace~$\udiruv$
by~$\udiruvinf$.  Here~$\detmetric$ is the determinant of~$\metric$.  This is
a ``constraint'' on the variation of~$\nugr_\udir$ along the unstable manifold
in chaotic flows and maps.  It was derived for arbitrary dimension in
Ref.~\cite{Thiffeault2002}.  It traces its origins in an analogous relation
for the stable manifold in Lagrangian coordinates, obtained using methods from
differential geometry~\cite{Tang1996,Thiffeault2001}.  A similar result was
obtained in the \twoD\ incompressible case in Ref.~\cite{Giona1998}, where it
was used to derive an invariant measure to characterize intermaterial contact
area.  In dimensions greater than two, the characteristic directions and
finite-time Lyapunov exponents associated with a chaotic flow must obey other
constraints~\cite{Thiffeault2002}, but~\eqref{eq:constraint} is the only one
we will use here.

Assuming that the material line has evolved long enough that the
constraint~\eqref{eq:constraint} is satisfied to any desired accuracy, and
assuming that the flow is incompressible, we rewrite the constraint as
\begin{equation}
	\frac{\pd}{\pd\Carclen}\log\nugr_\udir + \div\udiruvinf = 0,
	\label{eq:conslaw}
\end{equation}
where~$\Carclen$ is the arc length along the material line (\ie, the unstable
manifold).  Actually, incompressibility is sufficient to
have~\eqref{eq:conslaw}, but it is not necessary as any flow with~$\detmetric$
constant in space will satisfy~\eqref{eq:conslaw}.  This is the case, for
instance, in the Lorenz equations, where~$\div\velv=\text{constant}$.

Equation~\eqref{eq:conslaw} is a conservation law for the largest Lyapunov
exponent~$\lyap_\udir$: if neighboring unstable manifolds converge (diverge),
then the largest finite-time Lyapunov exponent increases (decreases).
Physically, the conservation law~\eqref{eq:conslaw} can be justified by
thinking of converging unstable manifolds (or equivalently the material line)
as ``squeezing'' fluid elements, thereby also stretching them if the flow is
incompressible.  In \threeD\ the unstable manifolds squeezes fluid elements by
the same amount because they must preferentially stretch along the unstable
manifold (the stretching that can occur along the intermediate
direction~$\ediruv_2$ is negligible, because we assumed
nondegenerate~$\nugr_\sigma$'s so that~\hbox{$\nugr_\udir\gg\nugr_2$} after
some time).

Returning to our model of a foliation of bends, the divergence of~$\udiruvinf$
evaluated on~$y=\fbend(x)$ is easily computed,
\begin{equation}
\div\udiruvinf \simeq \div\tanuv
	= -\frac{\fbend'\l(\fbend'' + \gbend_1(x,0)\r)}
	{(1 + {\fbend'}^2)^{3/2}}\,
	+ \frac{\gbend_2(x,0)}{(1 + {\fbend'}^2)^{1/2}}\,,
	\label{eq:divt}
\end{equation}
since the tangent~\eqref{eq:fbendtangent} to the curve is aligned with the
unstable manifold.  The subscripts~$1$ and~$2$ on~$\gbend$ denote
differentiation with respect to its first and second arguments.  We may think
of~$\gbend_1(x,0)$ as the variation in the tangent~$\tanuv$ along~$x$, and
of~$\gbend_2(x,0)$ as the variation along~$y$, both parametrized by~$x$, where
the~$x$--$y$ coordinates are defined in Fig.~\ref{fig:bendfoliation}.

We evaluate the other term of the constraint~\eqref{eq:conslaw}
on~$y=\fbend(x)$,
\begin{equation}
	\frac{\pd}{\pd\Carclen}\log\nugr_\udir
	= \udiruvinf\cdot\grad\log\nugr_\udir
	= \frac{1}{(1 + {\fbend'}^2)^{1/2}}\,
	\frac{d}{d x}\log\nugr_\udir\,,
	\label{eq:tdotgradLam}
\end{equation}
so that from~\eqref{eq:constraint}, \eqref{eq:divt},
and~\eqref{eq:tdotgradLam}, we find
\begin{equation}
	\frac{d}{d x}\log\nugr_\udir
	= \frac{\fbend'(\fbend'' + \gbend_2(x,0))}{1 + {\fbend'}^2}\,
	- \gbend_1(x,0).
\end{equation}
This can be integrated to yield
\begin{equation}
	\nugr_\udir
	= \aconst\,{(1 + {\fbend'}^2)}^{1/2}
	\exp\l(\int\frac{\fbend'(x)\,\gbend_1(x)}{1 + {\fbend'}^2}\dint x
	- \int\gbend_2(x,0)\dint x\r),
	\label{eq:nugrudirsoln}
\end{equation}
where~$\aconst$ is constant in space along the material line but depends on
time (it contains the exponential growth of~$\nugr_\udir$).  The integrals
in~\eqref{eq:nugrudirsoln} are indefinite.

First we show that we recover the result of Section~\ref{sec:simplebend} in
the limit~\hbox{$\gbend_1 = \gbend_2 = 0$} (a uniform foliation of bends).
To exhibit the relationship between stretching and curvature, we use the
expression
\begin{equation}
	\curvm(x) = \frac{|\fbend''(x)|}{(1+\fbend'(x)^2)^{3/2}}
	\label{eq:fbendcurvaturem}
\end{equation}
for the magnitude~$\curvm(x)$ of the curvature, and obtain
from~\eqref{eq:nugrudirsoln}
\begin{equation}
	\nugr_\udir
	= \aconst\,|\fbend''(x)|^{1/3}\,\curvm^{-1/3}
	\label{eq:onethirdlaw}
\end{equation}
after setting~\hbox{$\gbend_1 = \gbend_2 = 0$}.
Equation~\eqref{eq:onethirdlaw} agrees with~\eqref{eq:13law},
with~\hbox{$\aconst^3 = \beta^{-1}$}.  However, Eq.~\eqref{eq:onethirdlaw} has
been derived with no assumption as to the initial shape of the material line.

\begin{figure}
\psfrag{log k}{\raisebox{-1ex}{$\log\curvm$}}
\psfrag{log L}{\raisebox{1ex}{$\!\!\!\!\log\nugrpara_\udir$}}
\psfrag{k3}{$\curvm^{-1/3}$}
\includegraphics[width=.8\textwidth]{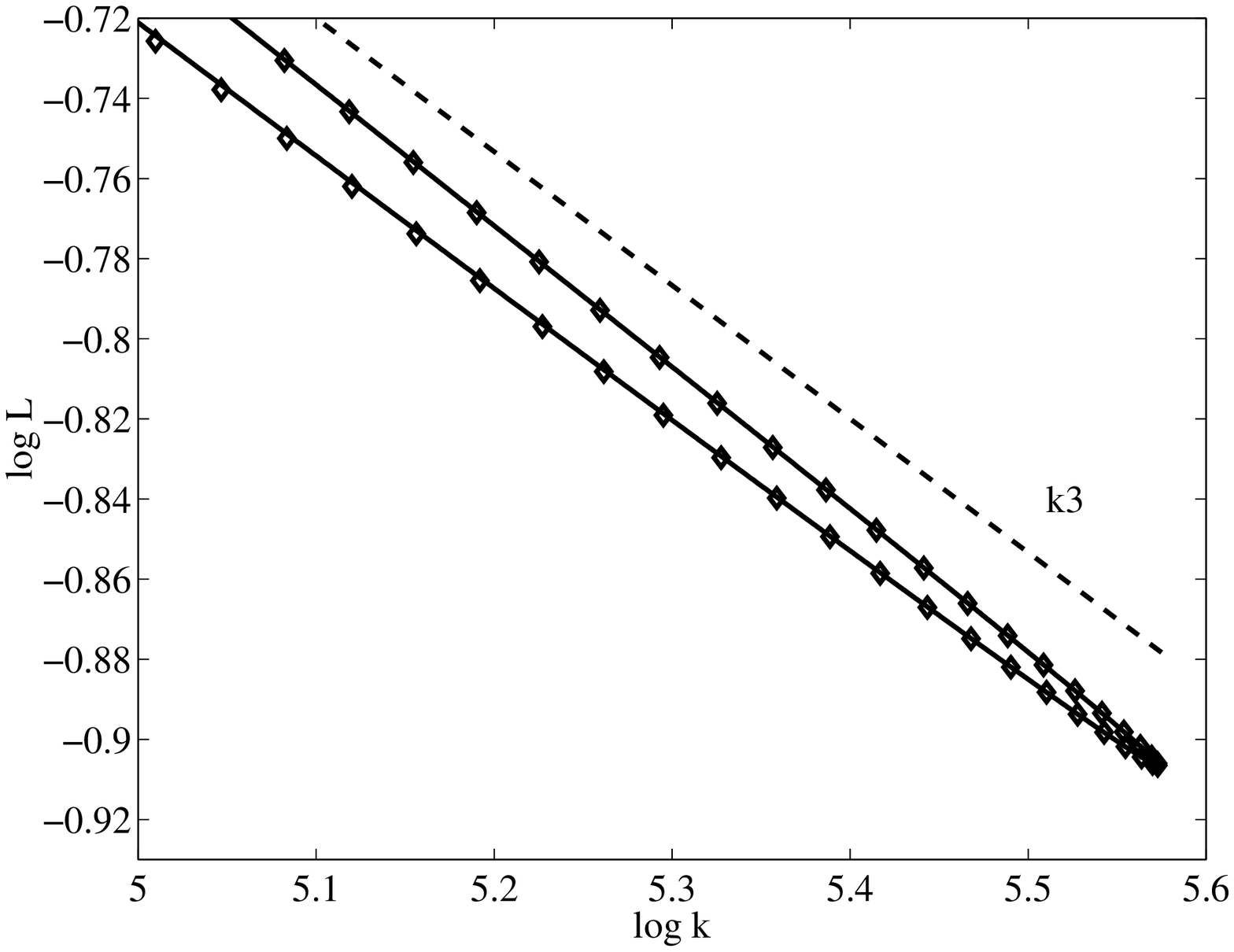}
\caption{Deviation from the~$-1/3$ law due to nonuniformity of bends.  The
  solid line is a fit using formula~\eqref{eq:lawdev}, and the diamonds are
  numerical data. The parameters used are \hbox{$\gbend_1/\curvm_0 = -0.01$},
  \hbox{$\gbend_2/\curvm_0 = 0.015$}, where~$\gbend_1$ controls the deviation
  from the~$-1/3$ law and~$\gbend_2$ the multi-valuedness of~$\nugrpara$.}
\label{fig:lawdev}
\end{figure}

Now we look for deviations from the~$-1/3$ law.  The integrals
in~\eqref{eq:nugrudirsoln} depend on the exact form of~$\fbend$
and~$\gbend_{1,2}(x,0)$.  But as a lowest-order approximation, near the tip of
the bend (at~$x=0$) we may assume that~$\fbend$ is quadratic
and~\hbox{$\gbend_{1,2}(x,0) \rdef \gbend_{1,2} = \mathrm{constant}$}, the
first nonzero terms in a power series.  The resulting stretching is
\begin{equation}
	\nugr_\udir
	= \aconst\,
	{\l(1 + {\fbend'}^2\r)}^{\tfrac{1}{2} + (\gbend_1/2\curvm_0)}
	\exp\l(-\gbend_2\, x\r),
	\label{eq:nugrudirsoln2}
\end{equation}
which may be rewritten in terms of the curvature~$\curvm(x)$
using~\eqref{eq:fbendcurvaturem},
\begin{equation}
	\nugr_\udir
	= \aconst\,\curvm^{-\tfrac{1}{3} - (\gbend_1/3\curvm_0)}
	\exp\l[\pm(\gbend_2/\curvm_0)\l((\curvm/\curvm_0)^{-2/3} - 1\r)\r].
	\label{eq:lawdev}
\end{equation}
Thus,~$\gbend_1$ appears as a deviation to the exponent of the~$-1/3$ law,
while~$\gbend_2$ gives an exponential, multi-valued correction (due to
the~$\pm$ sign).  For sharp bends,~\hbox{$\curvm_0\gg1$}, so that the
corrections due to~$\gbend$ are not visible.  In Fig.~\ref{fig:lawdev} we show
a bend where
the corrections are more important (\hbox{$\gbend_1/\curvm_0 = -0.01$},
\hbox{$\gbend_2/\curvm_0 = 0.015$}): the solid line shows that the fit
of~\eqref{eq:lawdev} to the numerical solution is very good near the tip,
where we expect it to be valid.

There are several sources of deviation from the~$-1/3$ law.  The first is from
the shape of the bend, as represented by the cubic and higher-order terms
in~$\fbend(x)$.  The second is in the variation of the tangent vector
field~\eqref{eq:fbendtangent}, that is, the shape of the neighboring bends, as
embodied by the function~$\gbend$ in~\eqref{eq:fbendtangent}.  In \threeD, the
variation in~$z$ (the direction perpendicular to the plane of the bend) must
also be taken into account, but we do not do so here.  What our analysis shows
is that the simplest configuration---a uniform foliation of bends---captures
the power-law behavior perfectly.  The leading-order corrections are
predominantly due to the function~$\gbend$, and not to the higher-order terms
of~$\fbend$, at least near the tip of the bend.  The reason for this is that
the~$\gbend$ terms enter~\eqref{eq:nugrudirsoln2} at lowest order in~$x$.

Numerical investigation indicates that the~$-1/3$ law often holds in \threeD\
around sharp bends, but not always---possibly a signature of torsion (the bend
is no longer planar).  A complete \threeD\ description would have to take into
account possible torsion in the material lines and will be the focus of future
research.

If the flow is compressible, then a simple modification
of~\eqref{eq:onethirdlaw} starting from~\eqref{eq:constraint} gives
\begin{equation}
	\nugr_\udir
	= \aconst\,\detmetric^{1/2}\,|\fbend''(x)|^{1/3}\,\curvm^{-1/3}\,,
	\label{eq:onethirdlawcomp}
\end{equation}
where we have assumed~$\detmetric$ depends only on~$x$.  The~$-1/3$ law will
thus not be affected by compressibility as long as the variations
in~$\detmetric$ are unimportant around sharp bends, which is usually the case
because of the localized nature of the bends.

We may consider instead of bends a foliation of horizontal lines, so
that~\hbox{$\fbend'(x)\equiv0$}.  Then we conclude
from~\eqref{eq:nugrudirsoln} that~\hbox{$\nugr_\udir = \aconst$},
for~\hbox{$\gbend_2(x,0)=0$}.  Such a foliation corresponds to a region of
vanishing curvature, and the constant stretching is reflected in
Fig.~\ref{fig:Lamt_vs_curv_stmap} (upper-left portion of the plot).  The
cellular flow, Fig.~\ref{fig:Lamt_vs_curv_crollsns}, does not exhibit the
constancy of~$\nugr_\udir$ in regions of low curvature, indicating that in
this less idealized situation the variation of the vector field~$\udiruvinf$
perpendicular to itself cannot be neglected.

\section{Conclusion}

Though stretching of fluid elements is more directly relevant to physical
problems such as mixing, understanding the kinematics of curvature is also
important.  A compelling reason is that, unlike stretching, curvature is
directly measurable from simple visualization experiments, being a purely
geometrical quantity.  Stretching and curvature are not independent, and it is
hoped that techniques such as those described herein can be used to relate
their distribution.  For instance, the~$-1/3$ law suggests that low values of
stretching are closely correlated to high curvature regions in a universal
manner, and thus the corresponding tail of the two distributions can have
similar properties.  There is some indication that this is the case, and
future work will address such statistical correlations.

\begin{acknowledgments}
The author thanks A. H. Boozer, D. Lazanja, and J. B. Keller for
helpful discussions.  This work was supported by the National Science
Foundation and the Department of Energy under a Partnership in Basic Plasma
Science grant, No.~DE-FG02-97ER54441.
\end{acknowledgments}


\end{document}